\documentclass{elsart}

\usepackage{graphicx}

\nocite{*}

\begin{document}

\begin{frontmatter}

\title{Temperature dependence of electric resistance and
magnetoresistance of pressed nanocomposites of multilayer nanotubes
with the structure of nested cones}

\author[M]{V.I. Tsebro\thanksref{e-mail}}
\author[M]{O.E. Omel'yanovskii},
\author[T]{E.F. Kukovitskii},
\author[T]{N.A. Sainov},
\author[K]{N.A. Kiselev},
\author[K]{D.N. Zakharov},
\address[M]{P.N. Lebedev Physics Institute,
Russian Academy of Sciences, 117924 Moscow, Russia}
\address[T]{Kazan Physicotechnical Institute, 420029 Kazan, Russia}
\address[K]{A.V. Shubnikov Institute of Crystallography, Russian
Academy of Sciences, 177333 Moscow, Russia}
\thanks[e-mail]{Author for correspondence (tsebro@sci.lebedev.ru).}

\begin{abstract}
Bulk samples of carbon multilayer nanotubes with the structure of
nested cones (fishbone structure) suitable for transport measurements,
were prepared by compressing under high pressure ($\sim$~25~kbar) a
nanotube precursor synthesized through thermal decomposition of
polyethylene catalyzed by nickel. The structure of the initial nanotube
material was studied using high-resolution transmission electron
microscopy. In the low-temperature range (4.2--100 K) the electric
resistance of the samples changes according to the law
$\ln \rho \propto (T_0/T)^{1/3}$,
where $T_0 \sim 7$~K. The measured magnetoresistance is quadratic
in the magnetic field and linear in the reciprocal temperature. The
measurements have been interpreted in terms of two-dimensional
variable-range hopping conductivity. It is suggested that the space
between the inside and outside walls of nanotubes acts as a
two-dimensional conducting medium. Estimates suggest a high value of
the density of electron states at the Fermi level of about
$5\times 10^{21}$~eV$^{-1}$cm$^{-3}$.
\end{abstract}

\end{frontmatter}

Investigations of electric transport properties of carbon nanotubes has
attracted great attention recently. According to theoretical
concepts~\cite{dress_book96}, an isolated nanotube can be either a
metal, or semimetal, or insulator, depending on such structural
parameters as its diameter, chirality, and the number of concentric
layers in it. Despite enormous difficulties in measurements of electric
parameters of isolated nanotubes or nanotube bundles, several attempts
undertaken recently have been
successful~\cite{langer96,dai96,ebbesen96}.
The latest published measurements~\cite{ebbesen96} clearly indicate the
presence of both metallic and insulating nanotubes in a single set of
samples prepared in the same conditions. The authors emphasized that
each multilayer nanotube manifested its specific conducting properties,
thus indicating a strong correlation between structural and electric
parameters.

In this connection, it is interesting to study, in addition to
the transport properties of isolated carbon nanotubes, the
conducting properties of bulk nanotube materials, in which
contacts between nanotubes and/or their sections are randomly
distributed. In our previous publication~\cite{tsebro051} we reported
on the conductivity temperature dependence and structure
(see also Ref.~\cite{tsebro050}) of carbon nanotube films fabricated by
evaporating graphite in an electron beam. The data of those
experiments were interpreted in terms of a three-dimensional
model of hopping conductivity with a Coulomb gap about
the Fermi level (the resistivity was described by the law
$\ln \rho \propto (T_0/T)^{1/2}$).
The density of states at the Fermi level for
films that contained, as shown by structural investigations,
mostly one-layer carbon nanotubes (isolated or assembled in
bundles) and had a relatively high conductivity was estimated
to be $g(\mu)\sim 10^{21}$~eV$^{-1}$cm$^{-3}$.
On the other hand,
films containing multilayer carbon nanotubes were characterized
by fairly large values of resistivity, which changed with
temperature to Mott's law,
$\ln \rho \propto (T_0/T)^{1/4}$.
In this case, estimates of the density of states,
$g(\mu)\sim 10^{18}$~eV$^{-1}$cm$^{-3}$, corresponded
to $g(\mu)$ for amorphous carbon. Amorphous carbon in significant
quantities was detected on the outside surfaces of multilayer nanotubes
in such films by electron microscopy~\cite{tsebro051,tsebro050}, and it
seems that the conductivity of such films can be attributed to the
presence of carbon.

It is well known that, in addition to one-layer and multilayer
nanotubes with walls made of coaxial carbon layers, there are nanotubes
whose walls consist of nested truncated cones (these are the so-called
fishbone-type structures~\cite{saito93}). Such nanocones are usually
detected at the ends of carbon nanotubes, but can also exist in the
form of independent objects among products of arc discharges in a
helium atmosphere~\cite{dress_book96b}, commonly used in synthesizing
carbon nanotubes.

In our recent work~\cite{kuk96,kis97,kuk97} we demonstrated that
thermal decomposition of polyethylene with nickel used as a catalyst is
a fairly efficient technique for fabrication of large quantities of
fishbone nanotubes. This technique allows one to manufacture in a
relatively short time considerable quantities (several grams) of fairly
homogeneous nanotube material. According to the data of thermal
analysis in oxiding atmosphere, the nickel content in this material is
less then 15\% by mass. Nickel is present in the material in the form
of nanoparticles, which can be eliminated completely by thermal
processing of the nanocomposite in vacuum at temperatures of up to
2800$^{\circ}$C~\cite{kis97}.

In this paper we present our measurements of electric
resistance versus temperature and magnetoresistance of bulk
nanocomposite samples fabricated by pressing the initial
powder of carbon fishbone nanotubes. The structure of the
carbon phase in the initial powder was imaged by a Philips
EM 430ST transmission electron microscope of high resolution
at an accelerating voltage of 200 kV. These measurements
demonstrated that the major part of the initial carbon
material was multilayer carbon nanotubes with lengths of
several micrometers, outside diameter of 40--50~nm, and internal
channel diameter of 9--20~nm. The tubes consisted of
almost rectilinear sections with lengths of 100--300~nm
turned with respect to one another. Figure~1 shows as an
example electron micrographs of the composite nanotube
material at (a) low, (b) medium, and (c) high resolution.

\begin{figure}[ht]
\begin{center}
(See separate attached jpg files)
\caption{Electron micrographs of nanotubes in the composite material at
(a) low, (b) intermediate, and (c) high resolution.}
\end{center}
\end{figure}

The analysis of micrographs indicated that the nanotube walls were
composed in most cases of 40--65 tapered graphite layers. The taper
angle varied along the tubes in the range of 16--35$^{\circ}$. The
inside diameter was also variable. The dimensions and shapes of wider
sections of the inside channel corresponded to those of catalytic
nickel nanoparticles, which were detected in most cases at the ends of
the nanotubes. We observed either so-called bamboo structures (with
taper angles of 20 to 25$^{\circ}$) or, more frequently, fishbone
structures with larger taper angles.

Bulk samples that could be used in transport measurements
were fabricated by cold pressing of nanotube powder
under high ($\sim$~25~kbar) pressure. Samples were shaped as
bars with dimensions of $\sim\ 1\times 2\times 3$~mm. Contacts for measuring
current and voltage across samples were made from a
conducting epoxy paste. Note that the samples were fairly
strong and their resistivity at room temperature was relatively
low: $\rho (300\,K) \sim 1\,\Omega $cm. The resistance was
measured as a function of temperature down to the liquid-helium
temperature in magnetic fields of up to 75~kOe.

In all samples under investigation, the resistance
changed with temperature most rapidly (about one order of
magnitude) in the temperature range between liquid helium
and $\sim\ $100~K, and the resistance followed the law
\begin{equation} \label{mott3}
R(T) = R_0 \ exp\,[(T_0/T)^{1/3}]\,,
\end{equation}
which is typical of variable-range hopping conductivity in
two dimensions. Figure~2 shows as an example two curves of
$\ln R$ vs. $T^{-1/3}$ plotted for samples Nos.~14 and 15.

\begin{figure}[ht]
\begin{center}
\includegraphics[width=12cm]{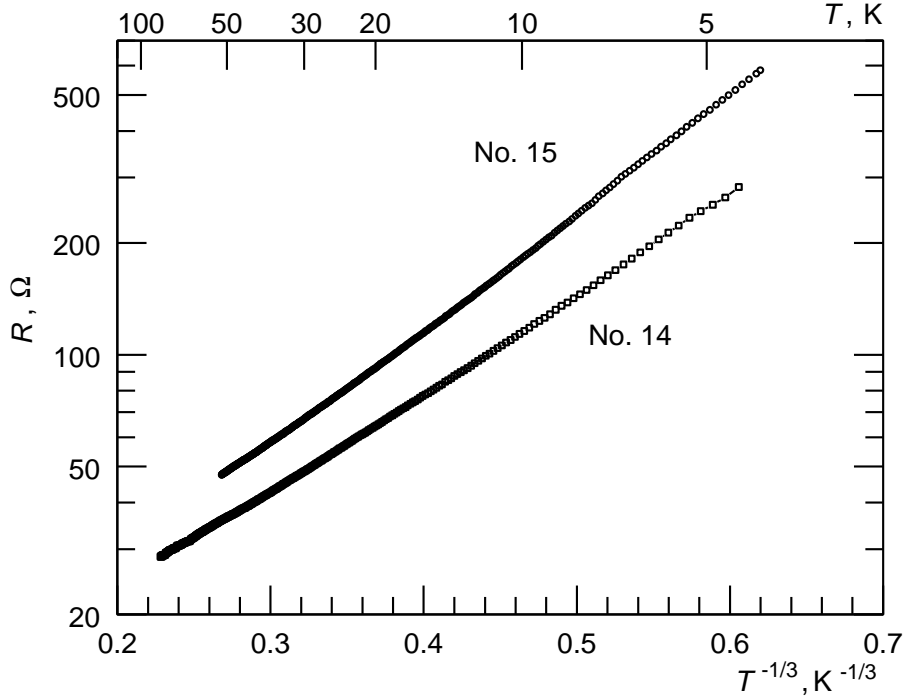}
\caption{Logarithmic resistance of samples Nos.~14 and 15 as a function
of $T^{-1/3}$.}
\end{center}
\end{figure}

It is known~\cite{shklov84} that in this case $T_0$  in Eq.~(\ref{mott3}) is
given by
\begin{equation} \label{t0mott3}
T_0 = \frac{13.8}{k_B \,g^*(\mu)\,a^2} \ ,
\end{equation}
where $g^*(\mu)$ is the two-dimensional density of states at the
Fermi level and $a$ is the localization length.

To the best of our knowledge, this is the first observation
of the dependence
$\ln R \propto (T_0/T)^{1/3}$ in a system with a relatively
low resistivity. Another interesting feature of our measurements
is low $T_0$ (for example, we found $T_0$ = 7,3~K in
sample No.~15, and in all tested samples $T_0$ was within the
interval of 6.5--7.5 K), which directly indicates, in accordance
with Eq.~(\ref{t0mott3}), that the density of states at the Fermi
level is high.

In this connection, it is of interest to measure the magnetoresistance,
especially as a function of temperature, since
these measurements would allow us to estimate directly the
localization length $a$ and then derive the two-dimensional
density of states $g^*(\mu)$ using Eq.~(\ref{t0mott3}).

It is known \cite{shklov84} that in systems with variable-range hopping
conductivity, the magnetoresistance is positive and (in
moderate magnetic fields) is given by the expression
\begin{equation} \label{mr}
\ln \left[ \frac{\rho (H)}{\rho (0)} \right]
= t\left(\frac{a}{\lambda}\right)^4
\left(\frac{T_0}{T}\right)^{3/p}
\equiv A(T) H^2 \ ,
\end{equation}
where $\lambda $ is the magnetic length, $t$ is a dimensionless factor
of about 0.0025, and $p = D + 1$ (where $D$ is the system
dimensionality).
Since $p$ = 3 holds in the case under consideration,
it follows from Eq. (\ref{mr}) that the magnetoresistance at a fixed
magnetic field should be inversely proportional to the temperature.

An example of magnetoresistance measurements versus
magnetic field at $T$ = 4.2 K for sample No.~15 is given in Fig.~3.
One can see that the magnetoresistance is adequately described by a
quadratic function of $H$ in the range of moderate magnetic fields,
$H <$30~kOe, and in higher magnetic fields it tends to a linear
function.

\begin{figure}[ht]
\begin{center}
\includegraphics[width=12cm]{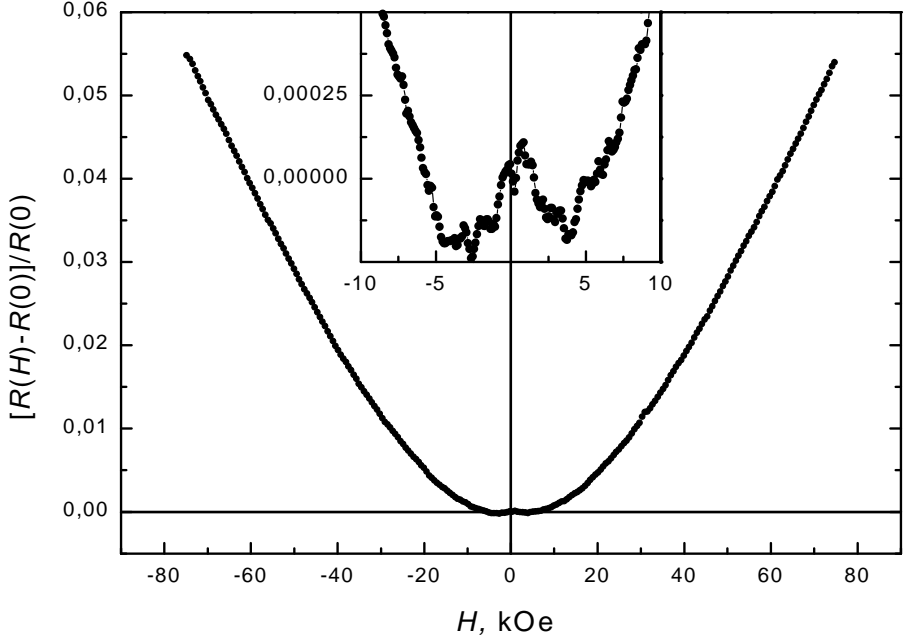}
\caption{Magnetoresistance of sample No.~15 versus magnetic field at
$T$ = 4.2~K. The inset shows the section of negative magnetoresistance
at low magnetic fields on an extended scale.}
\end{center}
\end{figure}

The behavior of magnetoresistance in low magnetic
fields is especially interesting. As a rule, the magnetoresistance
is negative on the section of the curve around zero and
becomes positive in fields higher than 7~kOe. As a result, we
have a small, broad region of negative resistivity at about
3--4~kOe. Moreover, several additional narrow local minima
(see the inset to Fig.~3) are observed superposed on this
broad peak. Note that the peaks in the inset to Fig.~3 are not
caused by noise, although their amplitudes are very small.
Experiments with repeated accumulation and averaging of
the signal dedicated to testing the reproducibility of such
measurements were performed (the results obtained by this
procedure are the ones plotted in the inset to Fig.~3), and
these experiments proved that the curves were reproducible,
even after warming the samples to the room temperature. It
seems that the negative magnetoresistance of the samples
and local minima are due to the discrete structure of the
conducting network formed by nanotubes. The broadest
minimum in the magnetoresistance at 3--4~kOe is tentatively
related to the average cell dimension in the network, and
local minima are ascribed to some additional characteristic
dimensions in the random network. When the applied magnetic
field reaches a value such that the magnetic flux
through a network cell equals the magnetic flux quantum
$hc/e$, the amplitude of the tunneling between nanotubes increases,
which causes a drop in the total resistance of the
system. A simple estimate yields a cell dimension of the
conducting network of about 120~nm at $H_{min}\sim 3.5$~kOe,
which seems plausible, given the structure of the nanotube
material shown by the electronic microscope.

The magnetoresistance of sample No.~15 as a function of
temperature under a magnetic field of 75~kOe is plotted in
Fig. 4 in terms of $\ln\,[R(H)/R(0)]$ and $T^{-1}$.
It is clear that the magnetoresistance at low temperatures is
reasonably well described by a linear function of $T^{-1}$, in
accordance with Eq.~(\ref{mr}). The localization length derived from
these measurements is $a$ = 17~nm. Thus, the two-dimensional density of
states at the Fermi level estimated using these data and
Eq.~(\ref{t0mott3}) is
$g^*(\mu)$ $\sim 7.5\times 10^{15}$~eV$^{-1}$cm$^{-2}$.

\begin{figure}[ht]
\begin{center}
\includegraphics[width=12cm]{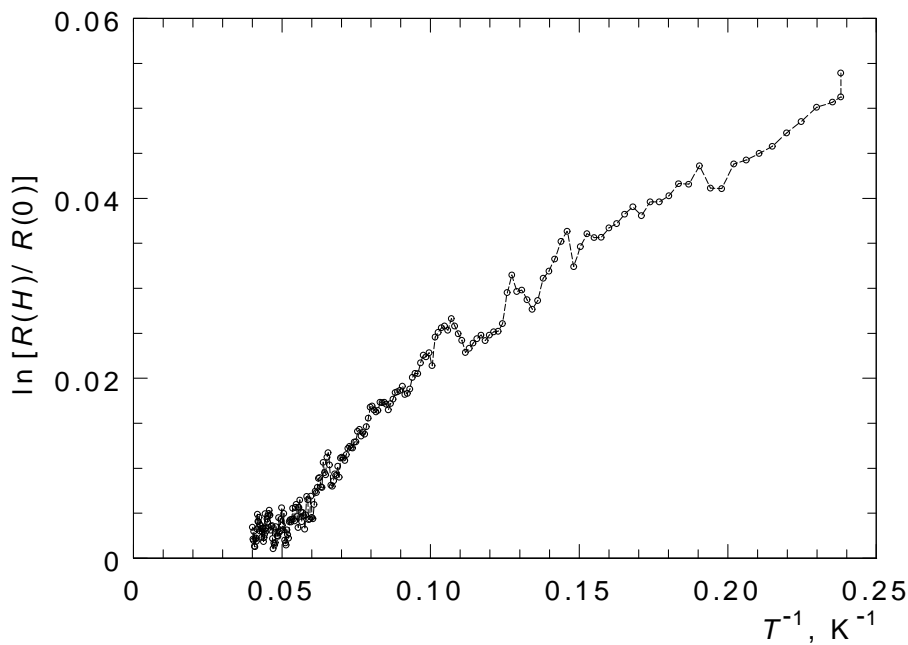}
\caption{Magnetoresistance of sample No.~15 versus temperature in a
magnetic field of 75~kOe plotted in coordinates
$\ln \,[R(H)/R(0)]$ and $T^{-1}$.}
\end{center}
\end{figure}

Assuming that the space between the inside and outside
walls of nanotubes acts as a two-dimensional medium, we
can estimate the three-dimensional density of states $g(\mu)$ at
the Fermi level. Using the relation $g^*(\mu) = g(\mu)\,d$, where $d$
is the average nanotube wall thickness (in this specific case it
is about 15~nm), we have
$g(\mu)$ $\sim 5\times 10^{21}$~eV$^{-1}$cm$^{-3}$.

It seems also interesting to estimate the two-dimensional
and three-dimensional densities, $n_S$ and $n_V$, of current carriers.
This can be done using the equation
\begin{equation} \label{ns}
n_S = 2g^*(\mu )\epsilon _0(T) ,
\end{equation}
where $\epsilon _0(T)$ is the energy band near the Fermi level containing
current carriers contributing to the hopping
conductivity ~\cite{shklov84}.
In the two-dimensional case, this band width
is given by the equation
\begin{equation} \label{eps0}
\epsilon _0(T) = \frac{(k_B T)^{2/3}}
{[g^*(\mu) a^2]^{1/3}} .
\end{equation}
At $T$ = 25~K we find from Eqs.~(\ref{ns}) and (\ref{eps0})
$n_S$ $\sim 9\times 10^{12}$~cm$^{-2}$,
$n_V$ $\sim 6\times 10^{19}~$cm$^{-3}$.

Thus, we have interpreted the low-temperature transport
measurements of pressed samples of randomly distributed
carbon nanotubes with a nested-cones structure in terms of
the two-dimensional variable-range hopping conductivity.
We have assumed that the space between the inside and outside
walls on nanotubes acts as a two-dimensional medium.
In our previous publication~\cite{tsebro051} the low-temperature properties
of carbon nanotubes were interpreted in terms of the three-dimensional
model of hopping conductivity with a Coulomb
gap in the density of states near the Fermi level. In both
cases, the resistance is described at low temperatures by the
law $\ln \rho \propto (T_0/T)^{1/n}$ with small $T_0$, which implies that these
carbon nanotube materials, with their various morphologies,
are characterized by very high densities of electron states at
the Fermi level of $\sim 10^{21}$~eV$^{-1}$cm$^{-3}$, which is a value
typical of metals. This result is important for understanding the
fundamental electronic properties of carbon nanotubes and
related materials and may also prove quite useful from the
viewpoint of practical applications.

The work was supported by the Russian Scientific Technological
Program {\em Topical Issues in Physics of Condensed
Media}, branch {\em Fullerenes and Atomic Clusters} (project No.~96147)
and International Center for Science and Technology (project No.~079).

Translation provided by Russian Editorial office.

\end{document}